\documentclass[aps,prl,preprint,groupedaddress]{revtex4-1}

\usepackage{graphicx}
\usepackage{dcolumn}
\usepackage{bm}
\newcommand{\la}{\left\langle}
\newcommand{\ra}{\right\rangle}
\newcommand{\PR}{Phys. Rev. }
\newcommand{\PRL}{Phys. Rev. Lett. }
\newcommand{\RMP}{Rev. Mod. Phys. }
\newcommand{\PL}{Phys. Lett. }
\newcommand{\JMMM}{J. Magn. Magn. Mater. }

\begin{document}


\title{Local electronic nematicity in the two-dimensional one-band Hubbard model}


\author{Kun Fang}
\affiliation{Department of Physics, University of Connecticut, Storrs, CT 06269, USA}
\author{G. W. Fernando}
\affiliation{Department of Physics, University of Connecticut, Storrs, CT 06269, USA}
\author{A. N. Kocharian}
\affiliation{Department of Physics, California State University, Los Angeles, CA 90032, USA}


\date{\today}

\begin{abstract}
Nematicity is a well known property of liquid crystals and has been recently discussed in the context of strongly interacting electrons. An electronic nematic phase has been seen by many experiments in certain strongly correlated materials, in particular, in the pseudogap phase generic to many hole-doped cuprate superconductors. Recent measurements in high $T_c$ superconductors has shown even if the lattice is perfectly rotationally symmetric, the ground state can still have strongly nematic local properties. Our study of the two-dimensional Hubbard model provides strong support of the recent experimental results on local rotational $C_4$ symmetry breaking. The variational cluster approach is used here to show the possibility of an electronic nematic state and the proximity of the underlying symmetry-breaking ground state within the Hubbard model. We identify this nematic phase in the overdoped region and show that the local nematicity decreases with increasing electron filling. Our results also indicate that strong Coulomb interaction may drive the nematic phase into a phase similar to the stripe structure. The calculated spin (magnetic) correlation function in momentum space shows the effects resulting from real-space nematicity.
\end{abstract}



\maketitle

\section{Introduction}~\label{intro}

Strongly correlated electron systems, such as the high temperature superconductors and other oxides, exhibit complex phase diagrams. In spite of intensive research during the past decade, there is no consensus on a possible mechanism for the superconductivity and certain new phases that are found from time to time. Recent measurements of magnetoresistivity, resistivity, IR reflectivity and neutron scattering in ruthenates\cite{ruth}, iron-based superconductors~\cite{Harriger,Chu} and cuprates~\cite{Ando,Hinkov,Tanatar,Daou,Fujita} show an interesting phase which locally breaks the rotational point group symmetry ($C_4$) along $x$ and $y$ axes (these directions for cuprates, defined along the Cu-O bond directions in the Cu-O plane) but preserves the translational symmetry. The symmetry of this phase greatly resembles the well-known nematic phase in isotropic liquid crystals, and hence the phase is called an electronic nematic state.

A two-dimensional (2d) electronic nematic phase is usually referred to as a phase that spontaneously breaks the symmetry of the underlying Hamiltonian associated with interchanging the $x$ and $y$ axes (defined in a certain plane) of the system~\cite{nematic}. Unlike conventional liquid crystals, in which the nematic state is due to the special shape of the molecules or the anisotropic interaction between molecules, the origin of an electronic nematic phase is the local electronic correlation. When $C_4$ symmetry is reduced to $C_2$ symmetry, the system is expected to have a $2$-fold degenerate ground state. The two ground states will be randomly distributed on the 2d plane and will form locally different domains similar to  magnetic domains in an Ising model. The effects of differently oriented domains are very likely to cancel out globally (i.e., on a larger length scale) and give rise to a lattice that appears uniform (i.e., without the above symmetry breaking).

It is technically difficult to detect the local electronic nematic symmetry breaking with disordered domains. Many experiments focus on $YBa_2Cu_3O_{6+x}$ because its orthorhombic structure plays a role similar to an external field that can align nematic domains along the same direction, so that it is easy to measure the macroscopic anisotropic effect. However the orthorhombicity also makes it hard to distinguish whether the anisotropic effect is caused by the distorted lattice structure or the local electronic correlation. Recent publications~\cite{Ando,Tanatar} have reported a clear and strong anisotropic resistivity even when the orthorhombicity is well suppressed and the temperature dependence of the anisotropic effect is different from that of a lattice distortion. This implies that the in-plane transport anisotropy cannot be explained solely by distorted lattice structures. This result was later supported by a cellular dynamical mean field theory (CDMFT)\cite{CDMFT} study of the two-dimensional Hubbard model by Okamoto {\it et al.}~\cite{Okamoto}. They showed that a strong nematic resistivity could be induced by a very small orthorhombic distortion as long as the interaction is strong enough to yield a Mott transition.

Another way to identify local nematicity is by using scanning tunneling spectroscopy (STS) with a rather high resolution. The STS can detect directly the electronic density of states within a nematic domain. In a recent publication, it is reported that the state-of-the-art STS measurements show  ``intra-unit-cell" electronic nematicity~\cite{Fujita} on a perfectly symmetric lattice along the $x$ and $y$ axes {\it i.e.,} a lattice that is not orthorhombic. Thus, this measurement provides the most direct evidence that the local nematic effect, in some strongly correlated materials, is purely electronic and originates from repulsive electron-electron interactions.

The experimental results from STS detection depict a fascinating picture of disordered local nematic states fluctuating throughout the undistorted cooper oxide plane, but, to our knowledge, there is no simple theoretical model supporting this idea. The motivation for this report is the possible existence of a ground state of the one-band Hubbard lattice Hamiltonian with spontaneous $x-y$ symmetry breaking. In order to identify and address such a locally nematic ground state, a real-space quantum cluster method~\cite{QCT} which can treat both the local and global effects at the same time is probably the best option for the problem. In this report, the variational cluster approximation (VCA)~\cite{VCA} calculations are performed on a $C_4$-symmetric, two-dimensional Hubbard square lattice. The results from the VCA calculations are used to make further analysis of this phase.

The report is organized as follows. The models and methods used in this paper, including a brief introduction to the VCA, are presented in Sec.~\ref{meth}. Sec.~\ref{res} first examines the existence of a local nematic phase and its coexistence with antiferromagnetism (Sec.~\ref{exist} and Sec.~\ref{af}). A local order parameter is introduced in Sec.~\ref{op} and the results on the dynamical structure factor are discussed in Sec.~\ref{dsf} while concluding remarks are given in Sec.~\ref{Summary}.

\section{Methodology}~\label{meth}

\subsection{Hamiltonian}~\label{Hami}

To explore a possible mechanism for local nematicity, we consider here the simplest one-band Hubbard Hamiltonian $\hat H$ on a square lattice

\begin{eqnarray}
{\hat H}=\sum\limits_{\la\bm r,\bm r'\ra\sigma}(-t{\hat c}^{+}_{\bm r\sigma}~{\hat c}_{\bm r'\sigma}+H.c.)+U \sum\limits_{\bm r} {\hat n}_{\bm r\uparrow}{\hat n}_{\bm r\downarrow},
\label{eqn:h}
\end{eqnarray}

where ${\hat c}_{\bm r\sigma}$ is the annihilation operator for electrons at site $\bm r$ with spin projection $\sigma$ and $U$ is the on-site screened Coulomb repulsion. $\la\bm r,\bm r'\ra$ denotes summation over the nearest neighbors. The parameter $t$ is the overlap integral between nearest neighbors. The two-dimensional square ($t_{\bm r,\bm r+x}=t_{\bm r,\bm r+y}=t$) lattice Hamiltonian is symmetric along $x$ and $y$ directions, {\it i.e.}, it has the underlying $C_4$ symmetry. All the energies reported here are measured in units of $t$.

\subsection{Variational Cluster Approximation (VCA)}~\label{vca}

There are many different numerical methods designed to solve the Hubbard model approximately. We will use the variational cluster approximation in this article. This is a quantum cluster~\cite{QCT} extension of Potthoff's self-energy functional approach (SFA)~\cite{SFA}. The SFA develops a variational process by using self-energy of a ``reference system" that can be exactly solved to obtain the approximate ground state properties of the ``real system" in the thermodynamic limit. In Ref.~\cite{SFA}, Potthoff proved that the grand potential $\Omega$ of a system, with one-particle parameter $\bm t$ and on-site interaction parameter $\bm U$, can be written as a functional of the self-energy $\bm \Sigma$ provided the system is not at the critical point of a phase transition~\cite{SFA}:
\begin{eqnarray}
&\Omega[\bm \Sigma]=Trln((\bm G_0^{-1}-\bm \Sigma)^{-1})+F[\bm \Sigma] \\
\nonumber\mbox{with \ \ \ } &\delta\Omega[\bm \Sigma]/\delta\bm\Sigma=0 \Leftrightarrow \bm G[\bm\Sigma]^{-1}=G_0[\bm\Sigma]^{-1}-\bm\Sigma
\label{eqn:sfa}
\end{eqnarray}
where $\bm G_0$ is non-interacting Green's function of the original model and $F[\Sigma(\bm t)]$ is the Legendre transform of the Luttinger-Ward functional~\cite{Luttinger}. The equations indicate that $\Omega[\bm \Sigma]$ is stationary at the exact self-energy and its value is the exact grand potential of the system. A ``reference system" is introduced with the same on-site interaction $\bm U$ but different one-particle parameter $\bm t'$, and the reference system can be exactly solved. In the VCA, the reference system consists of isolated, identical clusters from a decomposition of the original lattice. Because of the universality of the Luttinger-Ward functional with the same $\bm U$, the functional dependence of $F[\bm \Sigma(\bm t')]$ will be the same for the original system and the reference system. Therefore, $F[\bm \Sigma]$ can be extracted from the reference system and used in the original system, so that $F[\bm \Sigma]$ can be eliminated.
\begin{eqnarray}
\Omega_{\bm t}[\bm\Sigma]=\Omega_{\bm t'}[\bm\Sigma]-Trln(\bm G_0^{-1}-\bm\Sigma)+Trln(\bm G_0^{'-1}-\bm\Sigma),
\label{eqn:vca}
\end{eqnarray}
and the stationary point of the equation is at $\bm\Sigma=\bm\Sigma[\bm t]$. However, when evaluating the functional $\Omega_{\bm t}[\bm \Sigma]$ for the stationary point, one can only vary the functional within a subspace of self-energy $\Sigma(\bm t')$ instead of the full space $\Sigma(\bm t)$. This is where the approximation kicks in. By evaluating the stationary points of the functional in the restricted self-energy space, the grand potential and the Green's function of the original system can be found \emph{approximately}. It is possible that the approximation comes with multiple stationary points; it is also possible that some stationary points will shift or even disappear in a different self-energy space because of the approximation, but those not disappearing should be physically meaningful solutions and the true ground state resides at the state with the lowest grand potential~\cite{SFA}.

The VCA~\cite{VCA} is constructed upon the SFA by choosing the reference system an ensemble of independent, identical clusters from a decomposition of the original lattice. The cluster can be solved by either exact diagonalization~\cite{square} or Lanczos method~\cite{Lanczos}. The exact eigenvalues and eigenstates of the cluster lead to an accurate implementation of the correlation effect, at least within the dimensions of the cluster size, and long range correlation acts like a mean-field in this approach. Evaluating the stationary point(s) of $\Omega[\bm \Sigma]$ by varying the one particle hopping parameter in the cluster $\bm t'$ provides a good approximation to the exact ground state of the original system in the thermodynamic limit. The electronic density $n$ in the lattice can be controlled by the chemical potential of the lattice $\mu$. Due to the mean-field treatment beyond the cluster dimension, number of electrons per cluster can be non-integral. Hence, although the number of electrons in the reference system (the finite cluster) can only be an integer, it is indeed possible to achieve any electronic density from the VCA approach.

\subsection{Reference System for Solving Nematicity}~\label{ref}

\begin{figure}
\begin{center}
\includegraphics*[width=20pc]{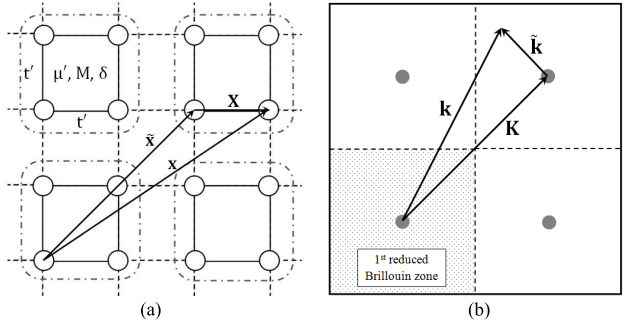}
\caption{(a) The reference system used in this study: the whole lattice is divided into identical $2\times 2$ (square symmetry) clusters with the same coupling $t'$ between the nearest atoms which is considered a variational parameter. Additional variational parameters are chemical potential $\mu^{'}$, effective Weiss field $M$ and fictitious deviation $\delta$ (see Sec.~\ref{ref} for details). (b) The full square is the first Brillouin zone (FBZ); shaded square denotes the reduced, first Brillouin zone (RFBZ) of the cluster superlattice. In a cluster representation, in the real lattice, upper-case letters represent lattice vectors within a cluster and the tilded letters denote a vector connecting the origins of two lattices; in the reciprocal lattice, we use upper-case letters for a vector between origins of two RFBZs and the tilded letters to represent the momentum vector within a RFBZ. For example, $\bm x =\bm X +\tilde{\bm x}$ in (a) and $\bm k =\bm K + \tilde{\bm k}$ in (b). We use the above definitions throughout this article.}
\label{fig:refer}
\end{center}
\end{figure}

In this report, we introduce  nematicity locally in the smallest cluster cell;  $2\times 2$  (square symmetry) clusters are used as the reference system (Fig.~\ref{fig:refer}(a)) to tile the infinite square and $t'$ is the hopping parameter in the reference system. Actually, one can also apply the larger-sized Betts clusters to reduce the size and edge effects~\cite{betts,JMMM}. The clusters use open boundary conditions according to Ref.~\cite{VCA} and $t'$ is a variational parameter for the VCA calculation.

Without coupling among the clusters, it is difficult to directly control the electronic density since the clusters have an integral number of electrons which is conserved. In the VCA, the chemical potential $\mu$ is varied to achieve different electronic density. However, the chemical potential of the clusters in the reference system, $\mu'$, should be different from $\mu$ in the original system, so $\mu'$ is also treated as a variational parameter. This treatment also ensures the thermodynamic consistency of the VCA, {\it i.e.,} the average electronic density $\la n\ra$ calculated from the trace of the Green's function, $Tr\bm G$, and the negative derivative of the grand potential $-\partial\Omega/\partial\mu$ are supposed to be the same~\cite{Aichhorn}.

The most significant improvement of the VCA over cluster perturbation theory~\cite{CPT} is that the VCA can describe the spontaneous symmetry-breaking orders by introducing a fictitious external field. In this paper, we will discuss two kinds of broken symmetries: antiferromagnetic (AF) and nematic. As for the AF order, one more term can be included in the reference system to represent the broken AF symmetry: $H_M\equiv M\sum\limits_{\bm r}e^{i\bm Q_{AF}\cdot \bm R}({\hat n}_{\bm r\uparrow}-{\hat n}_{\bm r\downarrow})$, where $\bm Q_{AF}$ is the AF wavevector and $M$ is the fictitious AF Weiss field whose value is determined by the variational procedure~\cite{Dahnken}. Here we only consider the simplest possibility for the AF case, which has a static AF order with $\bm Q_{AF}=(\pi,\pi)$. The nematic state gives rise to a fictitious difference between hopping along $x$ and $y$ directions; this difference can be addressed by introducing an additional term $H_\delta\equiv\delta\sum\limits_{\bm r}t'({\hat c}^+_{\bm r}{\hat c}_{\bm r+x}-{\hat c}^+_{\bm r}{\hat c}_{\bm r+y})+H.c.$, where $\delta$ denotes the fictitious deviation of the reference system from $C_4$ symmetry. The subscript ${\bm r+x(y)}$ indicates the nearest neighboring site along the $x(y)$ direction in the reference system. Here, we confine ourselves to $\delta>0$, which means that the hopping along the $x$ direction of the reference system is larger or equal to that along the $y$ direction. This will provide one ground state, and because the ground state for nematic order is two-fold degenerate, the other ground state is symmetric to the first ground state with a larger hopping along the $y$ direction. Both $M$ and $\delta$ are zero for the symmetric states and non-zero for the symmetry-breaking orders. It is important to note that $M$ and $\delta$ are both fictitious parameters, which indicate the influence of long-range order on the small clusters, so the bigger the size of the cluster, the smaller the fictitious parameters are. This rule can be used to justify the validity of the fictitious parameters: the case for $M$ is proved following this rule in Ref.~\cite{Senechal2}.

Finally, with these one-particle variational parameters, the reference system is represented by a Hamiltonian $\hat H'$ given by
\begin{eqnarray}
\nonumber{\hat H'}=\sum\limits_{\la\bm r,\bm r'\ra\sigma}(&-&t'{\hat c}^{+}_{\bm r\sigma}~{\hat c}_{\bm r'\sigma}+H.c.)-\mu'\sum\limits_{\bm r}{\hat n} \\
\nonumber &&+\delta\sum\limits_{\bm r}[t'({\hat c}^+_{\bm r}{\hat c}_{\bm r+x}-{\hat c}^+_{\bm r}{\hat c}_{\bm r+y})+H.c.]\\
&&+M\sum\limits_{\bm r}e^{i\bm Q_{AF}\cdot \bm R}({\hat n}_{\bm r\uparrow}-{\hat n}_{\bm r\downarrow})+U \sum\limits_{\bm r} {\hat n}_{\bm r\uparrow}{\hat n}_{\bm r\downarrow}
\label{eqn:ref}
\end{eqnarray}
where $\hat H'$ contains the same interacting term as $\hat H$ in Eq.~\ref{eqn:h} except for the modified one-particle terms which create a trial self-energy for the variational process. A series of trial self-energies are determined from the variational calculation which ends at the stationary points (which can be maxima, minima or saddle points) of the self-energy functional $\Omega[\bm\Sigma]$. The variational calculation involves four variational parameters ($t'$, $\mu'$, $M$ and $\delta$) making it difficult to find the correct stationary point(s), because it is not possible to search every corner of the variational space completely. Therefore, we carry out the search in this report as follows: (1) set $\delta=0$ and $M=0$ to find a trivial solution by the VCA; (2) vary only the fictitious deviation $\delta$ and the effective Weiss field parameter $M$ while fixing all the other parameters from the previous process; (3) as long as a non-trivial result ($\delta\neq 0$ or $M\neq 0$) shows up, an additional variational process with all the variational parameters will be performed to get the final result. The above three-step process is the most efficient way we have found in order to obtain a consistent variational behavior within the limited computer resources and time. We use a VCA program written by ourselves in FORTRAN. The program has been tested extensively and verified with results in some published work such as Refs.~\cite{VCA,Dahnken,Senechal2}.

\section{Results}~\label{res}
\subsection{Existence of Local Nematicity}~\label{exist}

The first question here is whether a local nematic state really exists in our simple model. The general experience from VCA calculations is that as long as the second step of the variational process mentioned in Sec.~\ref{ref} gives a nontrivial result, the true stationary point obtained in the third step is very likely to be nearby and the nontrivial result usually survives in the third step. Therefore, we can use results from the second step to test the validity of the model.

\begin{figure}
\begin{center}
\includegraphics*[width=20pc]{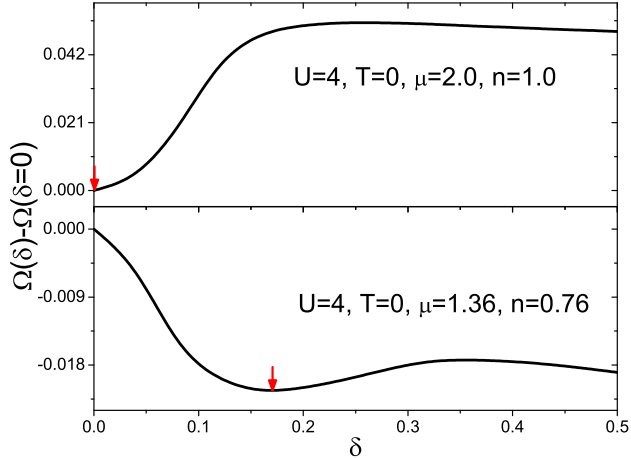}
\caption{Dependence of the grand potential $\Omega$ on the fictitious deviation $\delta$ is obtained from VCA calculation (the difference $\Omega(\delta)-\Omega(\delta=0)$ is plotted) on an infinite square lattice tiled by $2\times 2$ clusters (see Fig.~\ref{fig:refer}). The on-site Coulomb interaction is $U=4$ and temperature is $T=0$. The energy unit is given by the nearest neighbor hopping of the original lattice. The calculation is looking for the stationary (minimum or maximum) point of $\Omega(\delta)$ and the lowest stationary point is the best approximation to the ground state. The upper figure is calculated when chemical potential $\mu=2.00$ (half-filling). The lowest non distorted stationary point with $\delta=0$ implies that there is no nematicity for half-filling ground state. In contrast, the lower plot at $\mu=1.36$ (off half-filling, $n\approx 0.76$) has a proper stationary point with $\delta \neq 0$ shown by the red arrow at $\delta\approx 0.17$. The nonzero $\delta$ value suggests the nematic phase as a possible ground state solution in a single band Hubbard model. We further investigate $\delta$ up to $1$; in the region $0.36<\delta<1$ both plots are monotonically decreasing so there are no other stationary points (not shown in this figure).}
\label{fig:nemat}
\end{center}
\end{figure}

The dependence of the self-energy functional $\Omega$ on the fictitious deviation $\delta$ is calculated at different chemical potentials $\mu$ which determines the average electron density of the lattice by the thermodynamical equation: $\langle n\rangle =-\frac{1}{N}\frac{\partial \Omega}{\partial \mu}$ where $N$ denotes the number of lattice sites. Fig.~\ref{fig:nemat} is a plot of the difference $\Omega(\delta)-\Omega(\delta=0)$ as a function of $\delta$ at $U=4$ and $T=0$ for $\mu=2.0$ (half-filling) and $\mu=1.36$ (off half-filling, $n\approx 0.76$). Notice that parameter $\delta$ can be varied only within the interval [0,1]. There are several stationary points in the two plots but, generally, the lowest point is usually the best guess for the proper stationary point representing physical properties of the ground state. The red arrows mark the proper stationary points in both cases respectively. At $\mu=2.0$ (half-filling), the proper stationary point is at $\delta=0$ implying that a uniform self-energy is preferred, while at $\mu=1.36$ (far below half-filling), a non-zero fictitious deviation shows up which elucidates the possibility of a nematic state. Although it is not the exact stationary point of the whole variational space because other variational parameters are fixed, usually the true stationary point is not too far away, so the ground state at the true stationary point is very likely to remain nematic.

\begin{figure}
\begin{center}
\includegraphics*[width=20pc]{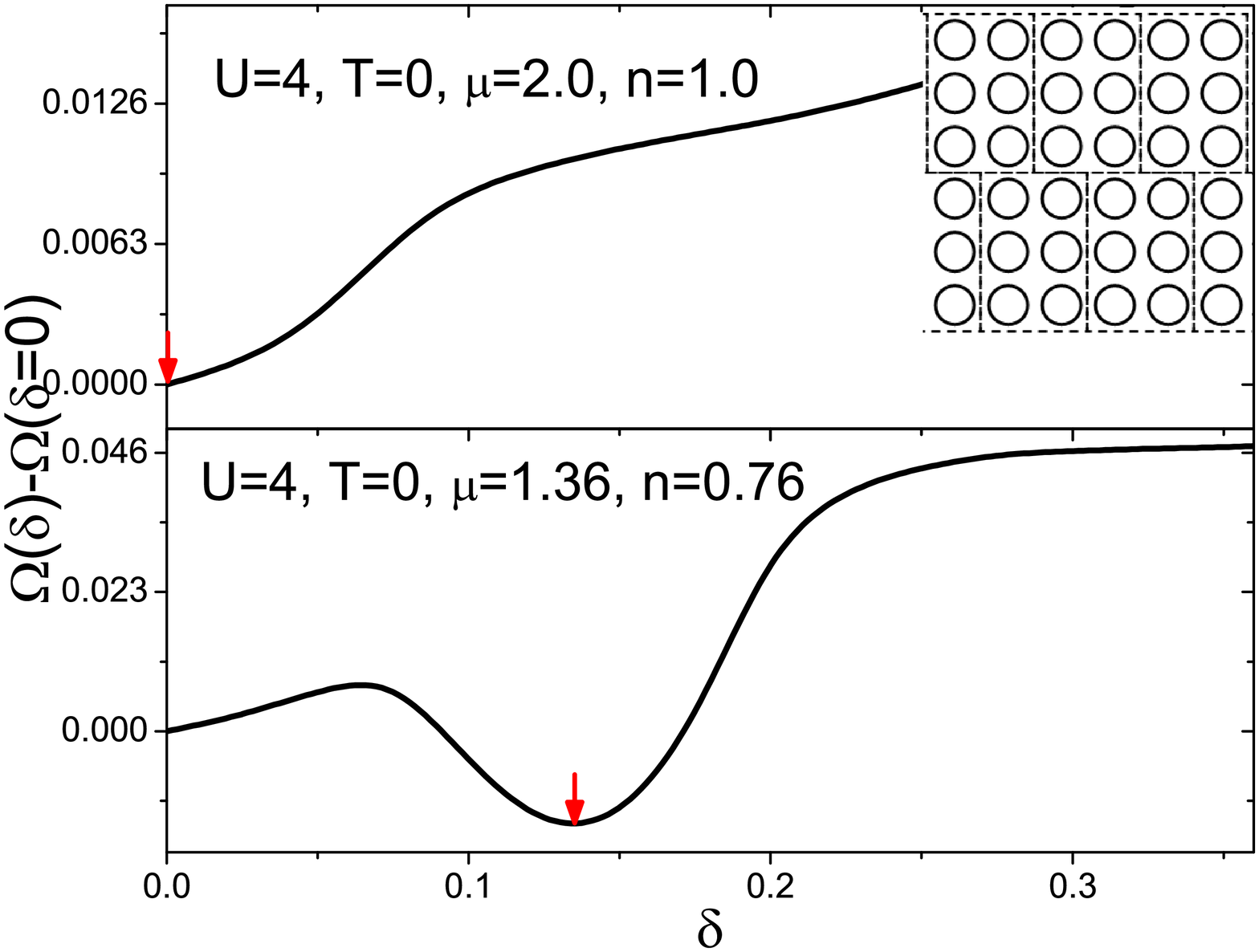}
\caption{$2\times 3$ clusters (the inset figure) are used as reference system. The fictitious deviation $\delta$ is obtained from VCA calculation. The difference $\Omega(\delta)-\Omega(\delta=0)$ is shown at $\mu=2.0$ ($n=1.00$) and $\mu=1.36$ ($n\approx 0.76$). At half-filling, the stationary point is still at $\delta=0$ indicating a normal state and a non-zero stationary point at $\delta=0.14$ is found at $\mu=1.36$. This shows the stability of the local nematic ground state in the pseudogap region at sufficient doping $n\approx 0.76$ and the smaller $\delta$ value is caused by the enlarged cluster size.}
\label{fig:clust6}
\end{center}
\end{figure}

We also use $2\times 3$ clusters as the reference system to justify the results related to nematicity. Fig.~\ref{fig:clust6} also shows a plot of $\Omega(\delta)-\Omega(\delta=0)$ as a function of $\delta$ and nonzero stationary point which demonstrates the existence of a nematic ground state of the original lattice. This figure is very similar to Fig.~\ref{fig:nemat} with stationary points at $\delta=0$ at half-filling and $\delta\neq 0$ at $n\approx 0.76$, but the nonzero fictitious deviation $\delta$ within this reference is smaller than that within the $2\times 2$ reference system. As discussed in Sec.~\ref{ref}, the parameter $\delta$ measures the residual effect of nematic symmetry breaking from other parts of the whole lattice, so the larger the cluster, the smaller the residual effect, {\it i.e.,} smaller $\delta$. Therefore, $\delta$ is expected to be smaller in the $2\times 3$ cluster-based calculation.

\subsection{Antiferromagnetism and Locally Nematic State}~\label{af}

There are many competing phases in strongly correlated materials; whether two given phases can coexist is also an issue very much tied to the correlations of these phases. Usually, order parameters of two phases are calculated at the same electronic density $n$: if both order parameters are nonzero at the same $n$, the two phases coexist. Here, in order to determine the possibility of the coexistence of AF order and nematic order, the effective Weiss field $M$ and the fictitious $\delta$ are both processed in the variational calculation as discussed in Sec.~\ref{ref}. We test the possibility of coexistence of the two states by comparing parameters $M$ and $\delta$. Although they are not order parameters because they are not physical observables, they can still provide a ``yes" or ``no" answer to whether the specific state exists at a certain electronic density $n$. Therefore, as long as $M$ and $\delta$ are nonzero at the same $n$, the AF order and nematic order can coexist.

\begin{figure}
\begin{center}
\includegraphics*[width=20pc]{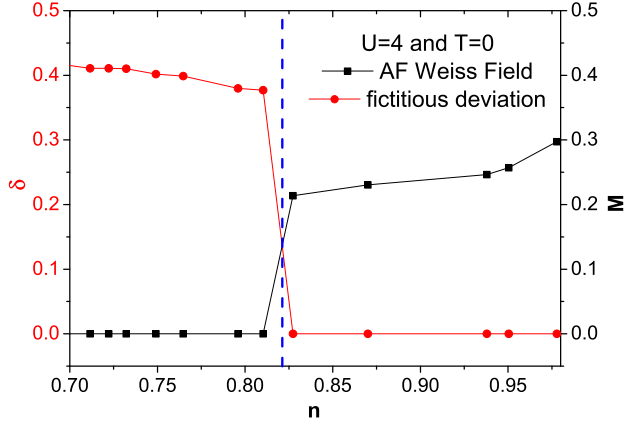}
\caption{The fictitious deviation $\delta$ and AF Weiss field $M$ are plotted simultaneously as a function of the electron density $n$ at $U=4$ and $T=0$. $M\neq 0$ means there is a static AF order with AF vector ${\vec Q}_{AF}=(\pi,\pi)$ and $\delta\neq 0$ indicates the nematic state appears, if they both are nonzero at the same $n$, that means the two states can coexist. In the plot, when $\delta\neq 0$, $M$ is zero and when $M\neq 0$, $\delta$ is always zero. This plot indicates that the static AF phase with
 ${\vec Q}_{AF}=(\pi,\pi)$ and the local nematic state cannot coexist.}
\label{fig:coex}
\end{center}
\end{figure}

After searching for possible stationary points in the variational space consisting of both nematic and AF phases, the fictitious deviation $\delta$ and the effective AF Weiss field $M$ are calculated simultaneously for different electron densities $n$ and the results are plotted in Fig.~\ref{fig:coex}. The plot clearly shows that when one variational parameter is nonzero, the other one turns out to be zero. This result is also verified by choosing an AF ground state as the starting point of the variational calculation. Fig.~\ref{fig:coex} indicates that the proper physical state of this simple Hubbard model cannot support the coexistence of static AF order and nematicity, i.e., a static AF order with an AF vector ${\vec Q}_{AF}=(\pi,\pi)$ cannot simultaneously break the local $C_4$ symmetry.

Experiments have shown that the nematicity can exist in the pseudogap region, where some local magnetism is still retained. First, it is possible that this magnetism is purely temperature driven. The reason we did not see such coexistence in the ground state is probably due to the fact that the static AF order with ${\vec Q}_{AF}=(\pi,\pi)$ requires an electronic structure with a perfect $C_4$ symmetry, while the nematic state breaks that symmetry. It is possible that the AF state coexists with local nematicity when it is not static or has a differently oriented AF vector. For example, a spin density wave phase or ${\vec Q}_{AF}=(\pi/2, 0)$, because this kind of AF phase does not strictly require the $C_4$ symmetry, the two phases may have a chance to coexist.

\subsection{Strongly Correlated Case}~\label{strong}

In order to study the case with strong electronic interactions, we consider the model at $U=8$. Although the fictitious deviation $\delta$ here is not a physical observable, it shows the response of the local electronic structure to nearby clusters within the nematic domain, so $\delta$, in some sense, represents the strength of the nematic order in the local nematic domain as long as it is calculated within the same cluster model.

\begin{figure}
\begin{center}
\includegraphics*[width=20pc]{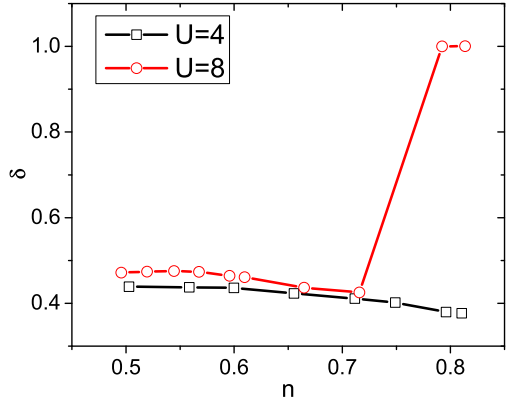}
\caption{The fictitious deviation $\delta$ from VCA calculation is plotted as a function of electronic density $n$ for $U=4$ and $8$. For $U=4$ (open square), $\delta$ is decreasing with electronic density up to $n<0.8$. For $U=8$ (open circle), $\delta$ behaves the same way as it does for $U=4$  within the overdoped region but increases up to about  one around optimal doping at $U=8$, which is probably a signal of the transformation into the stripe phase.}
\label{fig:nemn}
\end{center}
\end{figure}

In Fig.~\ref{fig:nemn}, values of $\delta$ at different electronic densities $n$ for $U=4$ and $U=8$ are shown. At $U=4$, the nematic state can exist from the overdoped region to the optimally doped region. The value of $\delta$ decreases with increasing electronic density, which implies that the nematicity is suppressed by local magnetic domains which begin to form as the electronic density becomes larger. At $U=8$, $\delta$ behaves the same way as that at $U=4$ in the overdoped region. However, around optimal doping, it jumps to about $\delta \approx 1$ at $U=8$. This abnormal behavior at $U=8$ points to a picture where the local hopping in one direction becomes close to zero, which suggests that the two-dimensional lattice is reduced to quasi one-dimensional chains. This strong coupling phenomenon is very similar to the stripe phase with charge and spin modulations coupled with broken rotational and translational symmetries found in some strongly correlated high $T_c$ superconducting materials. Such a strong trend to form inhomogeneous pattern of holes and spins is often referred to as charge and spin phase separation instabilities. The same abnormality was also reported by Okamoto {\it et al.}\cite{Okamoto}, where they have given similar explanations. However, in our opinion, the stripe phase simultaneously breaks both rotational symmetry and translational symmetry, but the VCA model does not examine the breaking of translational symmetry. In addition, as mentioned in Sec.~\ref{ref}, $\delta$ is an artificial field, so it is important to verify the tendency to form the stripe phase by monitoring relevant order parameters.

\subsection{Order Parameter of Local Nematicity}~\label{op}

In order to examine possible local nematicity, we look for a suitable order parameter. It is our belief that a cluster-based approach such as the VCA, where local correlations in the Hubbard model are treated exactly, is ideally suited for probing local order. Important insights can be obtained by examining a local order parameter which can be chosen as the difference between a physical observable along $x$ and $y$ directions such as the resistivity tensor~\cite{Tanatar}, spin correlation~\cite{Hinkov} or some other structure factor. For example, STS experimental work by Lawler et al.~\cite{Fujita} measured the local density of states (LDOS) within each ``sub-unit-cell" with a very high real-space resolution, which allowed them to successfully identify local nematicity. They defined a real-space order parameter to quantify real-space nematicity based on the LDOS at different atoms.

\begin{figure}
\begin{center}
\includegraphics*[width=20pc]{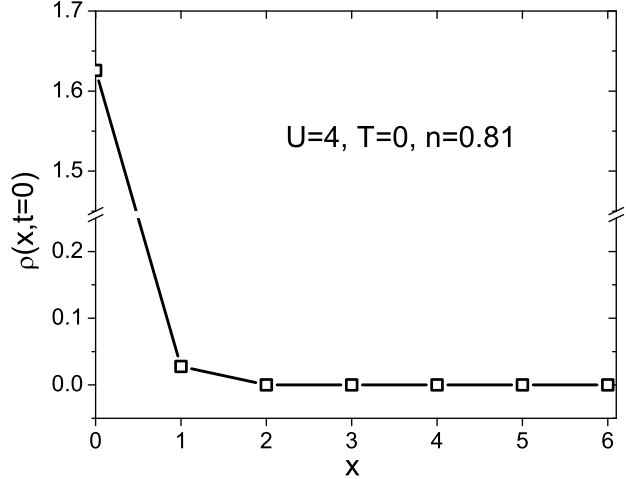}
\caption{The zero-time density correlation $\rho$ is calculated along the $x$ direction and plotted as a function of $x$, {\it i.e.,} $\rho=\rho(x,t=0)$. $\rho$ has a peak at $x=0$ and it quickly drops to a relatively smaller value as $x$ moves to $1$ lattice site away from the origin. When $x$ reaches $2$ lattice sites away from the origin, which is also out of the cluster scale, $\rho$ becomes almost $0$. As $x$ increases further, $\rho$ remains close to $0$. This plot shows that the zero-time density correlation is strongly localized in our model. The plots are calculated at $U=4$, $T=0$ and electronic density $n=0.81$ and a $10\times 10$ fixed mesh is used to sample the $\tilde k$-points.}
\label{fig:denscorr}
\end{center}
\end{figure}

Here, in searching for a locally nematic state in the two-dimensional Hubbard model, we define a nematic order parameter based on the zero-time, local density correlation function $\rho$ which is defined as
\begin{eqnarray}
\rho(\bm r,t=0)=\la {\hat n}(\bm r,t=0^+){\hat n}(\bm r'=\bm 0,t'=0)\ra-\la {\hat n(\bm r,t=0^+)}\ra^2
\label{eqn:dens}
\end{eqnarray}
where ${\hat n}(\bm r,t)$ is the electronic density operator at position $\bm r$ and time $t$, {\it i.e.,} ${\hat n}(\bm r,t)=\sum\limits_{\sigma} \la c^+_\sigma(\bm r,t+0^+)c_\sigma(\bm r,t)\ra$ with summation of different spins $\sigma$. The equation can be simplified by Wick's theorem
\begin{eqnarray}
\rho(\bm r,t=0)=G(\bm r,t=0^+)G(-\bm r, t=0^-)
\label{eqn:dens2}
\end{eqnarray}
where $G(\bm r,t)$ is the Green's function of the lattice at position $\bm r$ and time $t$. In order to calculate the Green's function and density correlation, Eq.~\ref{eqn:dens2} is rewritten using the cluster notations
\begin{eqnarray}
\rho(\bm r,t=0)=\frac{1}{(2\pi L)^2}\sum\limits_{\tilde k,\tilde k'}\int d\omega d\omega' e^{i(\tilde k-\tilde k')\tilde x}G(\bm X,\tilde k,\omega)G(-\bm X,\tilde k',\omega')
\label{eqn:densclst}
\end{eqnarray}
where $L$ is the number of $\tilde k$-points along a chosen direction in the calculation and $G(\bm X,\tilde k,\omega)$ is the cluster Green's function with notations defined in the Fig.~\ref{fig:refer}. We plot the position dependence of the zero-time density correlation along the $x$ direction in Fig.~\ref{fig:denscorr}. The figure shows that as $\bm r$ goes out of the cluster scale, the density correlation $\rho$ drops drastically, which indicates that $\rho$ is strongly localized within a small length scale. Except for the trivial correlation to itself, the most significant density correlation falls at the nearest neighbors. Let us denote the position of the nearest neighbor along the $x$ direction as $\hat x$ and that along the $y$ direction as $\hat y$. Then, the nematic order $O$ which breaks the $C_4$ rotational symmetry can be defined by the difference between density correlations of the nearest neighboring sites $\hat x$ and $\hat y$.

\begin{figure}
\begin{center}
\includegraphics*[width=20pc]{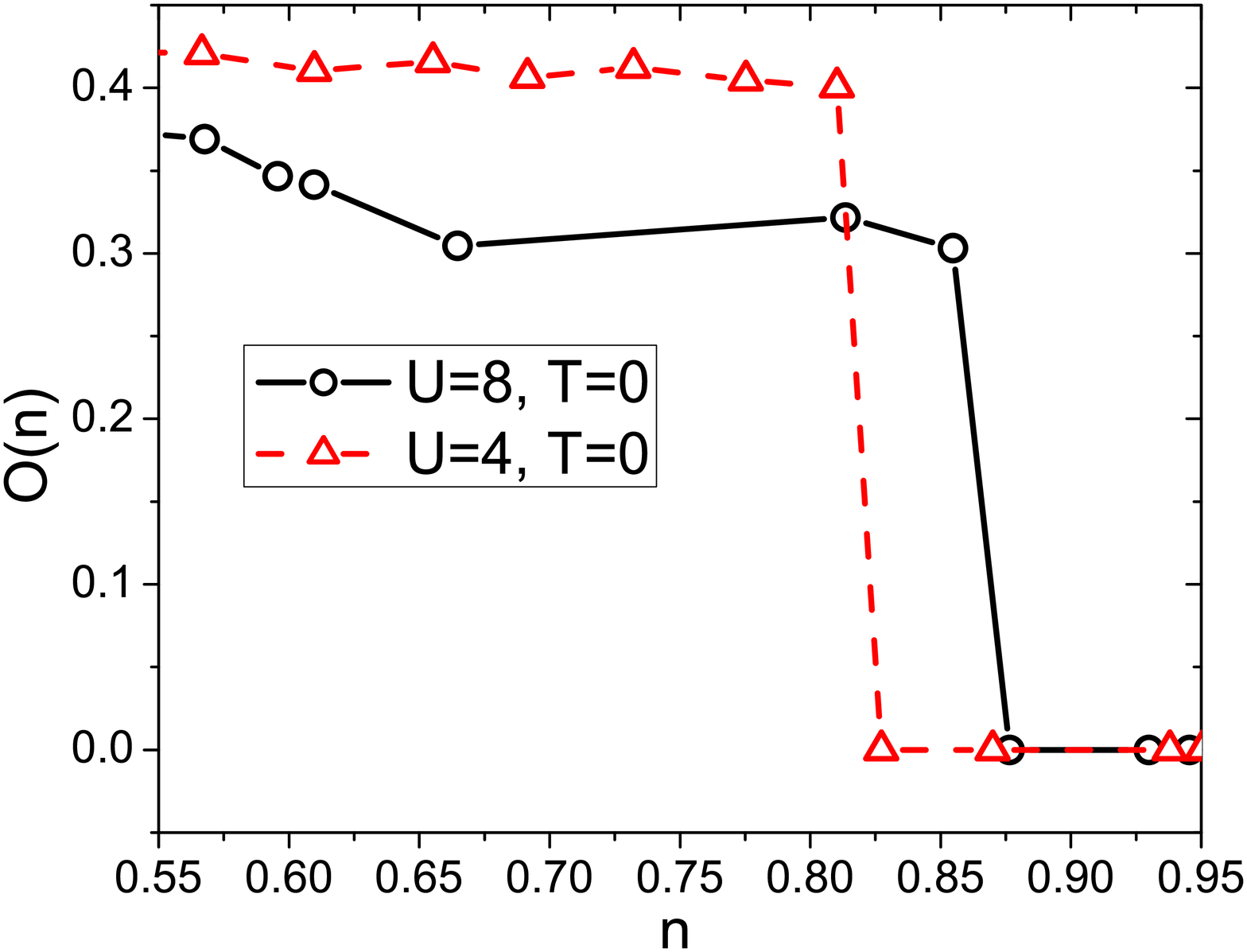}
\caption{Local nematic order parameter $O$ is plotted as a function of electronic density $n$ at $U=4$ (open squares) and $U=8$ (open circles). The temperature for both plots is $T=0$. This result is obtained by summing up to $10 \times 10$ momentum $\tilde k$-points in the cluster reciprocal lattice. $O$ is non-zero in the nematic phase with a broken $C_4$ symmetry and is zero in the phase with a symmetric order. The system undergoes a phase transition at about $n=0.82$ at $U=4$ and $n=0.86$ from a nematic phase to an AF phase. $O$ in the nematic phase does not change too much as the electronic density $n$ changes.}
\label{fig:op}
\end{center}
\end{figure}

\begin{eqnarray}
\nonumber O=\frac{\rho(\hat x,t=0)-\rho(\hat y,t=0)}{\rho(\hat x,t=0)+\rho(\hat y,t=0)}
\label{eqn:op}
\end{eqnarray}
In the normal (i.e., non-nematic) state, because of $C_4$ symmetry, the local correlation function, as defined above along the $x$ and $y$ is always zero. For a locally nematic state, the local correlation functions along the $x$ and $y$ directions will not be the same. Therefore, $O$ is zero in the normal state and nonzero in the locally nematic state. Fig.~\ref{fig:op} shows the evolution of the nematic order parameter $O$ with electronic density $n$ for $U=4$ and $8$. There is a clear phase transition between a nematic phase and a simple static AF phase. Since there is no evidence for coexistence between these two phases in our model (as discussed in the Sec.~\ref{af}), the figure shows clear phase boundaries, which are at about $n=0.82$ at $U=4$ and $n=0.86$ at $U=8$, between a $C_4$ symmetry phase and a broken $C_4$ symmetry phase correspondingly. The nematic order parameter $O$ in the nematic phase does not change too much with electronic density and indicates a relatively stable nematic phase. When compared with Fig.~\ref{fig:nemn}, where the fictitious deviation $\delta$ jumps to nearly $1$ at $U=8$ within the region $0.8<n<0.86$, $O$ does not show similar abnormal behavior. Although $\delta$ shows a tendency for the lattice to go into a stripe phase, the order parameter does not provide any support for the existence of this phase. In addition, the fictitious deviation field $\delta$ is an artificial field and therefore, the possible existence of a stripe phase in the lattice is still an open question.

The order parameter $O$ is defined using the density correlation function $\rho$ between nearest neighbor sites. As shown in Fig.~\ref{fig:denscorr}, The correlation function $\rho$ has a peak at the origin and decreases drastically at distance $1$, which suggests that the correlation function decays quickly as distance increases, so the electronic correlation is a short-range effect. Since electronic nematicity is caused by electronic correlation, the electronic nematicity is also a local effect. The value of $\rho$ drops to zero beyond the scale of the cluster due to the fact that the VCA is a mean field treatment beyond the cluster dimension. The mean field approximation suppresses fluctuations and hence the correlation goes to zero. The nematic ground states are two-fold degenerate with different orientations. In the symmetric lattice, the degenerate nematic ground states spread locally over the lattice. The randomly disordered local nematic orders with different orientations cancel out each other, so the whole lattice, at large length scale, is symmetric. Therefore, the nematic fluctuations in the uniform Hamiltonian can exist locally within each ``sub-unit-cell" and this conclusion gives strong support to the experimental results by Lawler et al.~\cite{Fujita}.

\subsection{Dynamical Spin correlation function}~\label{dsf}

Magnetic neutron scattering is a very useful tool to investigate magnetic and density fluctuations of a given material. The magnetic neutron scattering spectrum is related to the dynamical spin correlation function. It is defined as
\begin{eqnarray}
S(\bm r_i,\bm r_j,t)=\la\bm S(\bm r_i,t)\bm S(\bm r_j,0)\ra \\
\label{eqn:rscf}
S(\bm k,\omega)=\frac{1}{N}\sum_{i,j}\frac{1}{2\pi}\int_{-\infty}^{+\infty}dt e^{-i[\bm k(\bm r_j-\bm r_i)-\omega t]}\bm S(\bm r_i,\bm r_j,t)
\label{eqn:kscf}
\end{eqnarray}
The spin correlation of the 2D lattice includes longitudinal $S_\perp(\bm r_i,\bm r_j,t)$ ($S_\perp(\bm k,\omega)$) and transverse components $S_\parallel(\bm r_i,\bm r_j,t)$ ($S_\parallel(\bm k,\omega)$). The two components of real-space correlation functions can be written as
\begin{eqnarray}
S_\perp(\bm r_i,\bm r_j,t)&=&\la S_z(\bm r_i,t)S_z(\bm r_j,0)\ra \nonumber\\
&=&\sum_\alpha G_\alpha(\bm r_j,\bm r_i,-t)G_\alpha(\bm r_i,\bm r_j,t)
\label{eqn:scfperp}
\end{eqnarray}
\begin{eqnarray}
S_\parallel(\bm r_i,\bm r_j,t)&=&\la S_x(\bm r_i,t)S_x(\bm r_j,0)+S_y(\bm r_i,t)S_y(\bm r_j,0)\ra \nonumber\\
&=&\frac{1}{2}\la S^+(\bm r_i,t)S^-(\bm r_j,0)+S^-(\bm r_i,t)S^+(\bm r_j,0)\ra \nonumber\\
&=&\frac{1}{2}\sum_\alpha[G(\bm r_i,\bm r_j,t)+G(\bm r_j,\bm r_i,-t)]
\label{eqn:scfpara}
\end{eqnarray}where $G_\alpha$ is the real-space-time Green's function and $\alpha$ is the spin index (up or down). The longitudinal correlation function is the same as the result of density correlation function. The momentum-space correlation functions can be obtained by Fourior transform of the real-space correlation functions. Because we perform most of the calculations within the cluster framework, the DSF is finally written using cluster parameters defined in Fig.~\ref{fig:refer} as
\begin{eqnarray}
S_\perp(\bm k,\omega)=\int\frac{d\nu}{2\pi}\frac{1}{N}\sum_{RBZ}e^{-i\tilde{\bm q}\bm X_{ij}}&&\sum_{\bm X_{ij},\alpha}G_\alpha(\bm X_{ji},\tilde{\bm q},\nu) \nonumber\\
&& \times G_\alpha(\bm X_{ij},\bm k+\tilde{\bm q}-\bm Q,\nu+\omega)
\label{eqn:kscfperp}
\end{eqnarray}
\begin{eqnarray}
S_\parallel(\bm k,\omega)=\frac{1}{2}\sum_\alpha[G(\bm k,\omega)+G(-\bm k,-\omega)]&& \nonumber \\
=\frac{1}{2N}\sum_{\bm X_{ij},\alpha}[e^{-i\bm k\bm X_{ij}}G_\alpha(\bm X_{ij},\tilde{\bm k},\omega)+e^{i\bm k\bm X_{ij}}G_\alpha(\bm X_{ij},-\tilde{\bm k},-\omega)]&&
\label{eqn:kscfpara}
\end{eqnarray}
where $\bm X_{ij}$ is a lattice vector between the cluster sites $i$ and $j$; $\bm k$, $\tilde{\bm q}$ and $\bm Q$ are reciprocal lattice vectors as defined in the caption of Fig.~\ref{fig:refer}. $G$ is the Green's function in the cluster representation; $N$ denotes the number of sites in the lattice. 

$S_\perp(\bm k,\omega)$ represented by Eq.~\ref{eqn:kscfperp} involves coupling between two quasi-particles at $\tilde{\bm q}+\bm k$ and $\tilde{\bm q'}$ with energy $\omega+\nu$ and $\nu'$ respectively in $(\tilde{\bm k},\omega)$ space. The correlation spectrum is not zero only when $\tilde{\bm q}+\bm k=\tilde{\bm q'}+\bm Q$ and $\omega+\nu=\nu'$ are both satisfied, {\it i.e.,} $\nu(\bm k+\tilde{\bm q})-\nu(\tilde{\bm q})=\omega$. This condition corresponds to an electron-hole excitation, which is only valid around the Fermi surface ($\tilde{\bm k_F}$) of the cluster superlattice. $S_\parallel(\bm k,\omega)$ is the momentum-space green's function of the lattice.

\begin{figure}
\begin{center}
\includegraphics*[width=20pc]{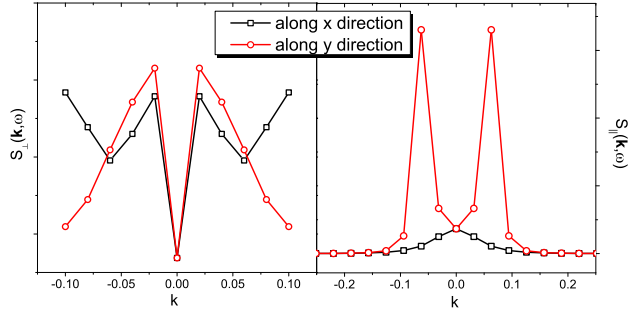}
\caption{The slice cut of the longitudinal and transverse spin correlation function $S_\perp(\bm k,\omega)$ and $S_\parallel(\bm k,\omega)$ along $(k_x,0)$ and $(0,k_y)$ at $U=4$ and $T=0$ with $\omega=0.001$ and electronic density $n=0.81$. The result of $S_\perp(\bm k,\omega)$ is obtained by summing up $20 \times 20$ momentum space points and integrating over the energy window from $-1.0$ to $1.0$. Both plots show large differences along the $x$ and $y$ directions, which elucidates that the nematic properties.}
\label{fig:dsf}
\end{center}
\end{figure}

The longitudinal and transverse spin correlation functions for magnetic excitations are calculated around in the momentum space at different energies. Fig.~\ref{fig:dsf} shows a slice cut of the DSF spectrum along $(k_x,0)$ and $(0,k_y)$. The large differences of the spectrum along $x$ and $y$ directions clearly show the asymmetry of the momentum-space density correlation function. This confirms that although the nematic state is a local effect in real-space, it still has a strong influence on some properties within the momentum space, especially on the two-particle correlations. 

Although our detailed theory only underlines the interplay of electron and spin density nematicities, the complete relationship between these order parameters and the high temperature superconductivity still needs further investigations. Strong competition of magnetism and superconductivity observed experimentally in both iron-pnictides and iron-chalcogenides implies that a complete description of nematic ordering must involve also the superconducting phase. The local electronic nematic ordering, which breaks $C_4$ symmetry, can be coupled to the introduced nematic order parameter $O$ by the fictitious parameter $\delta$ through the next-nearest or next-next-nearest neighbor couplings. This is one of the directions that can be pursued for studies of the extended Hubbard model in the future.

\section{Summary}~\label{Summary}
In conclusion, we have identified a local nematic phase away from half filling (in the overdoped region) by using the variational cluster approach to solve the one-band, two-dimensional Hubbard model. However, we do not find any coexistence of antiferromagnetic order and nematicity for their distinct symmetries. The variational cluster approximation shows evidence for electronic nematicity of the states consistent with the scanning tunnelling (spectroscopic-imaging) microscope measurements of the intra-unit-cell states in underdoped $Bi_2Sr_2CaCu_2O_{8+\delta}$ ~\cite{Fujita}. The breaking of rotational symmetry by the electronic structure within each $CuO_2$ unit cell becomes predominant in the pseudogap phase as the density of doped holes is reduced. Our results directly demonstrate that the nematicity can arise within a one-band isotropic (non-degenerate) Hubbard model. The nematic phase is suppressed as the electron density increases toward the half-filled state, but as the electron density gets close to optimal doping, while the Coulomb interaction is large enough, the nematicity suddenly gets enhanced and is likely to give rise to a phase similar to the stripe structure. The dynamical structure factor shows that the real-space local effects also influence the correlations in the momentum space and the asymmetry of the correlation pattern changes with different incident energies. This report is one of the first studies where a locally nematic state in the single band Hubbard square lattice is obtained without any additional input (such as next-nearest neighbor couplings, inter-site Coulomb interaction or multi-band structure).

\begin{acknowledgments}
The authors acknowledge A. V. Balatsky at Los Alamos National Laboratory for helpful discussions and the computing facilities provided by the Center for Integrated Nanotechnologies, a U.S. Department of Energy, Office of Basic Energy Sciences user facility at Los Alamos National Laboratory (Contract DE-AC52-06NA25396) and Sandia National Laboratories (Contract DE-AC04-94AL85000). This work was performed also, in part, at the Center for Functional Nanomaterials, Brookhaven National Laboratory, which is supported by the U.S. Department of Energy, Office of Basic Energy Sciences, under Contract No.DE-AC02-98CH10886.
\end{acknowledgments}

\end{document}